\documentclass[conference]{IEEEtran}
\IEEEoverridecommandlockouts
\usepackage{cite}
\usepackage{amsmath}
\usepackage[cmintegrals]{newtxmath}
\usepackage{algorithmic}
\usepackage{graphicx}
\usepackage{textcomp}
\usepackage{xcolor}
\usepackage{booktabs}
\usepackage{tikz}
\usetikzlibrary{positioning,arrows.meta,shapes.geometric,calc,decorations.pathreplacing}

\usepackage[hidelinks]{hyperref}

\def\BibTeX{{\rm B\kern-.05em{\sc i\kern-.025em b}\kern-.08em
    T\kern-.1667em\lower.7ex\hbox{E}\kern-.125emX}}

\newcommand\copyrighttext{%
  \scriptsize
  \noindent\textbf{Accepted to IEEE AIxHEART 2026.} 
  \textcopyright~2026 IEEE. Personal use of this material is permitted. 
  Permission from IEEE must be obtained for all other uses, in any current or future media, 
  including reprinting/republishing this material for advertising or promotional purposes, 
  creating new collective works, for resale or redistribution to servers or lists, 
  or reuse of any copyrighted component of this work in other works.%
}

\newcommand\copyrightnotice{%
  \begin{tikzpicture}[remember picture,overlay]
    \node[anchor=south,yshift=10pt] at (current page.south) {%
      \fbox{\parbox{\dimexpr\textwidth-2\fboxsep-2\fboxrule\relax}{\copyrighttext}}%
    };
  \end{tikzpicture}%
}

\begin{document}

\title{MAIA: Multi-Agent Intent Articulation for Requirement Discovery in Art Commissions}

\newif\ifdoubleblind
\doubleblindfalse 

\ifdoubleblind
  \hypersetup{
    pdftitle={MAIA: Multi-Agent Intent Articulation for Requirement Discovery in Art Commissions},
    pdfauthor={},
    pdfkeywords={Digital art, Human computer interaction, Multi-agent systems, AI-mediated communication}}
\else
  \hypersetup{
    pdftitle={MAIA: Multi-Agent Intent Articulation for Requirement Discovery in Art Commissions},
    pdfauthor={Yu-Chao Wang, Yanhong Lu, Yingjie Victor Chen, and Tim McGraw},
    pdfkeywords={Digital art, Human computer interaction, Multi-agent systems, AI-mediated communication}}
\fi

\ifdoubleblind
\author{\IEEEauthorblockN{Anonymous Author(s)}
\IEEEauthorblockA{Anonymous Affiliation}}
\else
\author{\IEEEauthorblockN{Yu-Chao Wang, Yanhong Lu, Yingjie Victor Chen, and Tim McGraw}
\IEEEauthorblockA{\textit{School of Applied and Creative Computing} \\
\textit{Purdue University}\\
West Lafayette, Indiana, USA \\
\{wang3368, lu699, chen489, tmcgraw\}@purdue.edu}
}
\fi

\maketitle
\copyrightnotice

\begin{abstract}
In bespoke art commissions, laypeople know what they feel but lack the words to specify it: one participant wanted a laid-off truck driver depicted as ``a ghost in his own machine'' but left the medium, scale, and palette unsaid. We frame this as an articulation bottleneck at an under-served upstream stage: requirement discovery, which precedes any artist or image generator and forces the commissioner to constitute intent in the first place. We present MAIA (Multi-Agent Intent Articulation), a multi-agent system that scaffolds this stage through Socratic inquiry under a ``Verification over Invention'' rule, turning vague affect into a text-only brief of visual terms the user verifies. In a within-subjects study ($N{=}16$), the full configuration produced a large, significant gain in Cognitive Support over a minimal baseline ($r{=}0.97$, $p_{\mathrm{FDR}}{=}0.003$; LMM $p_{\mathrm{FDR}}{<}0.001$). Thematic analysis traces the same mechanism, and a validator gate structurally blocks unratified content. An exploratory blind review by three professional concept artists provides complementary descriptive evidence that the rewritten briefs were more complete and executable.
\end{abstract}


\section{Introduction}

\noindent A commission is decided before the artist is hired: laypeople arrive with proto-ideas but lack the domain terminology to express them \cite{belkin1980anomalous}, so they either omit the visual parameters that matter or bury them in emotional prose, and the artist is left to reconstruct what was meant. We call this failure an \textit{articulation bottleneck}: the gap between a felt proto-idea and an actionable visual code.

This bottleneck sits at a stage prior work has largely bypassed: \textit{requirement discovery}, the upstream transition from an ambiguous impulse to a concrete concept \cite{fang_generative_2025}. Generative-AI tools have expanded how laypeople explore visual concepts \cite{Verheijden_Funk_2023, Jiang_Wu_Deng_Long_Tang_Li_Liu_Jin_Zhang_Qi_2024, Fu_Newman_Going_Feng_Lee_2025}, and AI-Mediated Communication (AI-MC) \cite{Hancock_Naaman_Levy_2020} has polished how messages land. Yet both assume the intent behind a request already exists and is stable. In a commission it does not: the user's task is to \emph{constitute} that intent in the first place, and a system that treats half-formed affect as a finished vision papers over the gaps with hallucination.

\begin{figure*}[t]
    \centering
    \includegraphics[width=\linewidth]{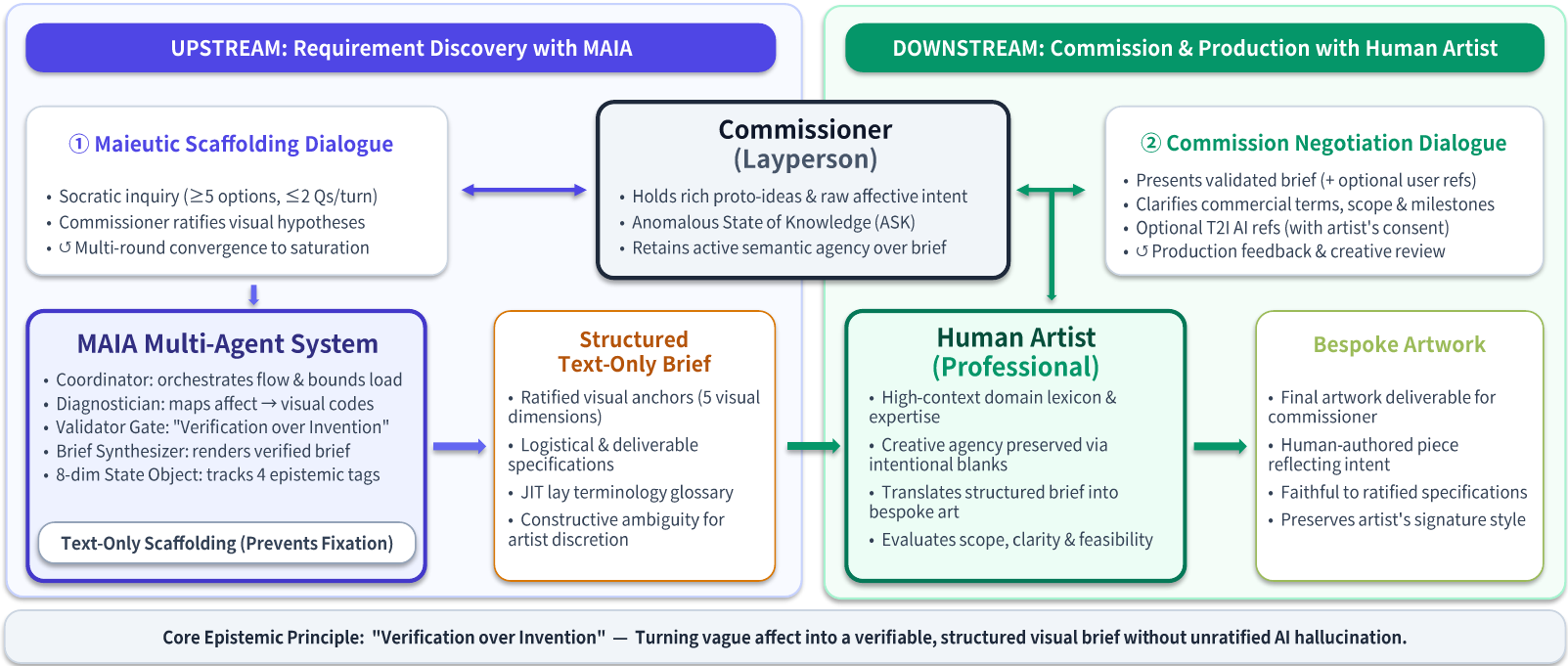}
    \caption{The commissioner--MAIA--artist pipeline (stage \textcircled{1} evaluated in this study). \textbf{(\textcircled{1})} Maieutic Scaffolding Dialogue (upstream, text-only): Socratic inquiry turns the commissioner's proto-ideas into a structured \emph{text-only brief}, with every AI-proposed visual element ratified under ``Verification over Invention.'' \textbf{(\textcircled{2})} Commission Negotiation Dialogue (downstream): the validated brief reaches a professional artist, who clarifies commercial terms and produces the artwork. The pipeline is text-only: MiniMax M2.5 neither generates nor reads images.}
    \label{fig:position}
\end{figure*}

To bridge this gap we introduce MAIA (\textbf{M}ulti-\textbf{A}gent \textbf{I}ntent \textbf{A}rticulation), a multi-agent active co-articulator. Through iterative Socratic inquiry it turns low-context impulses into concrete visual terms the user can verify, giving commissioner and artist a shared reference. MAIA operates \emph{upstream} and \emph{text-only} (Fig.~\ref{fig:position}): exposing laypeople to high-fidelity, AI-hallucinated imagery too early risks overwriting the latent intent the brief is meant to surface. Its guiding rule, ``Verification over Invention,'' requires that any visual element the AI proposes that the user did not supply be ratified before entering the brief.

We contribute:

\begin{enumerate}
    \item A conceptual framing of the expert--layperson gap as an articulation bottleneck at a distinct upstream stage, requirement discovery, where intent must be constituted rather than optimized.
    \item A multi-agent mechanism, MAIA, that operationalizes this through Socratic inquiry under ``Verification over Invention,'' with a validator gate and a four-state epistemic tag that make the system's commitments inspectable.
    \item Initial evidence ($N{=}16$) that this scaffolding substantially improves Cognitive Support during requirement discovery: the gain holds across both statistical tests and the qualitative findings, with the validator gate structurally blocking unratified content. An exploratory artist-side blind review (three professional concept artists, sampled briefs) adds initial evidence that the AI-rewritten briefs are more executable.
\end{enumerate}

\section{Related Work}\label{sec:related}

Prior work spans expert--layperson communication, generative AI in design, and AI-mediated communication; together, these threads locate the under-served stage of requirement discovery.

\subsection{Expert-Layperson Communication}

Digital art commissions are transient, low-context engagements between laypeople and professionals separated by a wide cross-domain knowledge gap \cite{liu2025exploring}, where conveying underlying context matters more than merely transmitting information \cite{eppler2007knowledge}. This friction is rooted in the expert--layperson \textit{contextual lexicon} gap. Artists hold a high-context lexicon yet suffer the ``Curse of Knowledge,'' assuming clients share their internal visualization of technical terms \cite{camerer:1989:ck, cao2020expertise}. Laypeople, conversely, operate in an \textit{anomalous state of knowledge} (ASK) \cite{belkin1980anomalous}, with strong affective intent but no terminology to articulate it, and unreliable reporters of their own knowledge gaps, which hierarchy discourages them from clarifying \cite{liu2025exploring, song2025personalized}.

To compensate, laypeople fall back on verbosity, packing far more structure into a message than an expert formulation would \cite{cao2020expertise}. Prior work treats this as information overload to be pared down \cite{eppler2004overload, bischof2011caring}, but doing so mischaracterizes the creative process: these verbose narratives are valuable, if untailored, proto-ideas. The friction is therefore an articulation bottleneck: proto-ideas fail to become actionable codes. Resolving it means turning affect into concrete technical requirements while leaving room for \textit{constructive ambiguity}. The system should fix the visual anchors that matter without forcing arbitrary choices the user never made, so the brief stays open to the artist's judgment, much as a good architect fixes the budget and site constraints but leaves the building's form to the designer.

\subsection{Generative AI in Design}\label{subsec:genai}

Existing generative-AI work in design concentrates on downstream artifact production and workflow efficiency \cite{Fu_Newman_Going_Feng_Lee_2025, Inie_Falk_Tanimoto_2023, Shi_Gao_Jiao_Cao_2023}, leaving underexplored requirement discovery, the upstream transition from ambiguous impulses to concrete concepts \cite{fang_generative_2025}. This overlooks the layperson's articulation bottleneck. Experts suffer \textit{design fixation} \cite{jansson_design_1991} but can still traverse the problem space; a layperson often cannot get started at all, which is why a static brief form that assumes a complete specification does not help them. Text-to-image generation does not resolve this either: exposing laypeople to AI-hallucinated imagery risks overwriting latent intent \cite{pearson2019human}, a visual fixation analogous to design fixation.

An AI mediator fits this stage better than an early conversation with the artist would. It lowers the social cost of voicing half-formed ideas \cite{shneiderman2007creativity} and keeps the iteration loop fast \cite{csikszentmihalyi1990flow}; it also keeps the artist out of the loop early on, so the artist is not anchored by an ambiguous first brief \cite{tversky1974judgment}. Existing scaffolding tools do not quite fill this role. \textit{CausalMapper} \cite{huang_causalmapper_2023} leans toward formal causal reasoning, whereas the \textit{Ideation Compass} \cite{valk_ideation_2023} stays close to open-ended reflection; neither negotiates between artistic freedom and the technical constraints a commission requires.

\subsection{AI-MC in Creative Workflows}

AI-MC, defined by Hancock, Naaman, and Levy as interpersonal communication in which an AI modifies, augments, or generates text \cite{Hancock_Naaman_Levy_2020}, has become a measurable research program, largely within the same group. In a referential task, Mieczkowski et al.\ found smart replies made messages more positive without senders absorbing that tone, while partners rated smart-reply users as less socially attractive \cite{Mieczkowski_Hancock_Naaman_Jung_Hohenstein_2021}; a follow-up traces how such effects redistribute agency and communicative roles \cite{Mieczkowski_Hancock_2022}. Meng et al.\ add that in emotional support, blind raters judged messages most authentic under an \textit{AI-guided} mode, where the writer takes the AI's direction and then drafts personally, because personal experience stayed in the text \cite{Meng_Zhang_Qin_Lee_Lee_2025}. The premise is that the sender already knows what to say. A commission breaks this: composing the message is itself how one figures out what to say, so a translator that treats half-formed affect as complete hallucinates the missing detail and drifts to the model's defaults, leaving commissioner and artist with the illusion of mutual understanding.

Creative workflows therefore move from passive message optimization to \textit{active epistemic co-articulation}, centered on \emph{maieutic inquiry}: probing latent intent via Socratic questioning \cite{chang_framework_2024}. This is distinct from ``maieutic prompting'' \cite{jung_maieutic_2022}, a model-side recursive self-consistency technique. We operationalize this as three design goals in Section~\ref{sec:system}, recasting common ground as a short list of fixed technical requirements with interpretive room for the artist.

A further gap is dyadic: most prior work studies the creator's or generator's side, whereas the commissioner--artist pair, and the \emph{upstream} moment before they ever speak, has received little attention and almost none from the artist's perspective. Our study addresses the commissioner's side of this pair and complements it with an exploratory blind review in which professional concept artists rate the executability of the resulting briefs (Section~\ref{sec:artist_review}).

\section{System Design}\label{sec:system}

Based on the epistemic gaps identified in Section~\ref{sec:related}, we propose MAIA, a multi-agent epistemic scaffolding system that helps the user traverse the problem space, so that user and artist reach a shared understanding of the brief. We implemented MAIA on OpenCode (OpenCode Desktop v1.1.60; all agents use MiniMax M2.5 at temperature $t=0.2$) as one user-facing coordinator that dispatches to specialized sub-agents. This harness enforces role separation architecturally.

\subsection{Design Goals}

To operationalize the role of the AI as a co-articulator, we established three design goals:

\subsubsection{DG1: Fostering Crystallization}

The first goal is to turn vague intent into the kind of specificity an artist can work from. We do not want a mechanical translation that swaps lay words for jargon, because that loses what the user actually meant. Instead the system connects a low-context phrase to the visual code behind it, so the artist spends less effort decoding the brief while the user keeps control of what it means.

\subsubsection{DG2: Epistemic Scaffolding via Maieutic Inquiry}

Because a layperson's first input rarely traverses the full problem space, the system does not wait for a complete specification. Drawing on the Socratic method, it offers either/or choices, so that vague impulses become explicit decisions before the brief reaches the artist, much as a human expert walks a client out of their initial assumptions \cite{paton2011brief}. Bounded options are a deliberate choice. Unrestricted freedom is itself a barrier for a novice who does not know where to begin; constraints make it easier to start \cite{stokes_creativity_2006}. People also discover what they want by reacting to alternatives, since generating a preference from a blank slate is hard \cite{slovic_construction_1995}; options let a commissioner evaluate a proposal far faster than produce one from scratch. To keep this scaffolding from becoming a straitjacket, every question always offers an ``Other'' escape that lets the user type a free-form answer, preserving their authorship over anything the system failed to anticipate.

\subsubsection{DG3: Intent Stewardship through Verification}

To mitigate the ``semantic drift'' of unconstrained large language model (LLM) generation, the system adheres to ``Verification over Invention'': any visual element the AI proposes that is not explicitly present in the user's input is treated as a tentative hypothesis and must be ratified by the user before inclusion.

\subsection{System Architecture}

A standard chat agent runs everything through one monolithic prompt that mixes analysis, checking, and synthesis together. MAIA instead splits these jobs apart, giving each to a dedicated agent with its own instructions, permissions, and knowledge. The interaction is no longer a one-shot command: the system alternates between diagnosing what the user means and scaffolding the next question, with the epistemic state written down at each step. A persistent knowledge base (a per-turn working file plus an append-only archive) carries this state and the operating principles across turns.

\subsubsection{Coordinator and Specialized Roles}

The \textbf{Maieutic Coordinator} is the system's sole user-facing agent, and it is \emph{forbidden} to reason about visual content on its own: on every turn it must delegate analysis to specialized sub-agents \emph{before} it may reply. These roles separate concerns that a monolithic chat agent would conflate. An \textbf{Intent Diagnostician} maps each affective descriptor to candidate visual codes and supplies a lay \emph{Just-in-Time} (JIT) definition for any technical term it introduces, returning only a compressed handoff (Fig.~\ref{fig:pipeline}). A \textbf{Response Validator} then independently audits that handoff and returns only \textsc{Approved} or \textsc{Revision-Needed}; it never authors user-facing content. A \textbf{Brief Synthesizer} renders a fully validated state into a cohesive art-director's mandate. This separation is what lets the validator gate be structurally independent of the generation it audits.

\subsubsection{State Object and Four-State Epistemic Tagging}

The mechanism that operationalizes both DG3 and constructive ambiguity is a shared \textit{State Object} tracking eight dimensions: five visual (Subject Definition, Composition, Lighting, Color Palette, Artistic Style) and three logistical (Usage/Purpose, Dimensions/Format, Deliverable Specs). Each dimension is continuously tagged with one of four epistemic states: \textsc{Unaddressed} (no information yet), \textsc{Inferred} (an AI hypothesis that \emph{must} be ratified before use), \textsc{Explicit} (stated or confirmed by the user), or \textsc{Discretion} (deferred to the artist). This tagging makes the system's epistemic commitments inspectable: an \textsc{Inferred} tag is by construction a hypothesis, and a \textsc{Discretion} tag is protected from re-interrogation.

\subsubsection{The Maieutic Turn: A Four-Step Pipeline}

\begin{figure}[!t]
    \centering
    \begin{tikzpicture}[
        node distance=2mm,
        every node/.style={font=\footnotesize},
        flowstep/.style={draw, rounded corners, align=center, text width=40mm, inner sep=2pt, minimum height=5mm, fill=gray!8},
        agt/.style={draw, rounded corners, align=center, text width=40mm, inner sep=2pt, minimum height=5mm, fill=orange!12},
        gate/.style={draw, rounded corners, align=center, text width=40mm, inner sep=2pt, minimum height=5mm, fill=yellow!22, very thick},
        arr/.style={-Stealth, thick},
        lbl/.style={font=\scriptsize}
    ]
        \node[flowstep] (in) {User message};
        \node[flowstep,below=of in] (s1) {\textbf{1. Log \& Distill}\\\scriptsize Coordinator};
        \node[agt, below=of s1] (s2) {\textbf{2. Diagnose}\\\scriptsize @intent-diagnostician\\\scriptsize CoT: Extract-Map, Diagnose,\\\scriptsize Hypothesize+JIT, Strategize};
        \node[gate, below=of s2] (s3) {\textbf{3. Validate}\\\scriptsize @response-validator};
        \node[flowstep,below=of s3] (s4) {\textbf{4. Respond}\\\scriptsize multiple-choice ($\geq$5 options, $\leq$2 Qs)\\\scriptsize update State Object};
        \node[flowstep,below=of s4, fill=green!10] (out) {Reply to user};
        \draw[arr] (in)--(s1);
        \draw[arr] (s1)--(s2);
        \draw[arr] (s2)--(s3);
        \draw[arr] (s3)-- node[right,lbl]{APPROVED} (s4);
        \draw[arr] (s4)--(out);
        \draw[arr] (s3.west) -- ++(-9mm,0) |- (s2.west) node[pos=0.28,left,lbl]{REVISION};
    \end{tikzpicture}
    \caption{The serial per-turn pipeline: the coordinator cannot reply until the validator returns \textsc{Approved}, and a \textsc{Revision-Needed} verdict loops back to re-diagnosis; the final brief is re-audited under a stricter regime.}
    \label{fig:pipeline}
\end{figure}
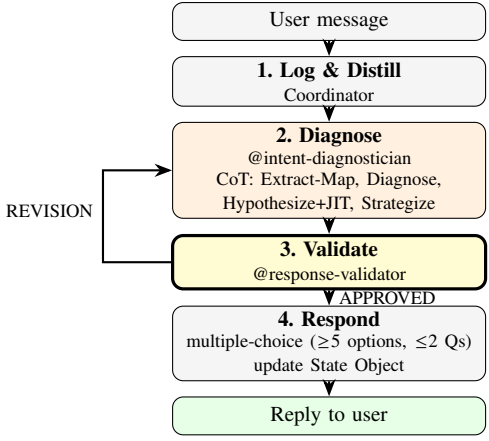

Every user turn executes a serial four-step pipeline (Fig.~\ref{fig:pipeline}); out-of-order or parallel invocation triggers a rollback. \textbf{(1) Log \& Distill}: the coordinator records the user's raw input and a distilled intent. \textbf{(2) Diagnose}: the coordinator invokes the Intent Diagnostician, which runs a chain-of-thought (CoT) analysis, appends it to the shared log, and returns a compressed handoff. \textbf{(3) Validate}: the coordinator forwards the handoff to the Response Validator, which returns \textsc{Approved} or \textsc{Revision-Needed}; the coordinator may not proceed until \textsc{Approved}. \textbf{(4) Respond}: the coordinator blends creative inquiry with logistics, updates the State Object, and renders every open question as a multiple-choice prompt (at least five annotated options, at most two substantive questions per turn to bound cognitive load).

\subsubsection{Convergence and Brief Finalization}

A commission reaches \textit{Semantic Saturation} when all five visual dimensions are tagged \textsc{Explicit} or \textsc{Discretion}, leaving no \textsc{Unaddressed} or unresolved \textsc{Inferred} items. The Brief Synthesizer then drafts the brief, which the validator re-audits under a stricter regime (zero \textsc{Inferred} items) before delivery.

\section{Methodology}\label{sec:methodology}

\subsection{Participants and Design}
\label{sec:design}
We recruited $N=16$ laypeople (Age $M=23.81$, $\text{SD}=2.81$) without formal visual arts training, to preserve ecological validity for ASK in artistic articulation; the study was approved by the Purdue University Institutional Review Board (IRB; STUDY2026-00000157), and all participants provided informed consent. We employed a 2$\times$2 within-subjects design crossing System Condition (\textit{Baseline LLM} vs.\ \textit{Maieutic Scaffolding}) with Scenario Type (\textit{Narrative-Affective} vs.\ \textit{Commercial-Promotional}), with condition order counterbalanced via a balanced Latin square. Each participant encountered each system and scenario exactly once, in opposite pairings, so the System and Scenario main effects are within-subject while the System$\times$Scenario interaction is carried between subjects. Both conditions used the same OpenCode Desktop and the text-only MiniMax M2.5 model, with identical agent harness settings. The Baseline was OpenCode's default single-agent chat, instructed to help refine the brief, whereas the Maieutic condition added the full multi-agent architecture of Section~\ref{sec:system}.

\subsection{Stimuli and Procedure}
\label{sec:procedure}
We designed two textual stimuli lacking explicit visual directives: a \textit{Narrative-Affective} task (a book illustration for a character facing an existential crisis) and a \textit{Commercial-Promotional} task (club callout art with specified mascot traits); the AI was blind to the task it served. In each block, participants first drafted an initial brief ($B_{\mathrm{base}}$), produced a final brief ($B_{\mathrm{final}}$) with their assigned system, and completed post-task evaluations. A 5-minute distractor task and demographic survey served as washout before the second block.

\subsection{Artist-Side Blind Review (Exploratory)}
\label{sec:artist_review}

To examine the artist's side of the commissioner--artist pair, we recruited three professional concept artists for an exploratory blind review under the study's IRB approval. We sampled the briefs of participants P01--P04: 8 tasks (4 Narrative, 4 Commercial), 16 briefs total, with 4 tasks under each system condition; all three artists rated the same 16 briefs (48 ratings, fully crossed). On 7-point scales they rated five Visual Specification Completeness (VSC) sub-dimensions (lighting, composition, color palette, artistic style, subject definition), executability (whether the brief could be executed without further clarification), artist autonomy, and brief professionalism. We report inter-rater reliability (ICC) and describe the task-level gains; because the eight rated tasks are clustered within four participants, each of whom contributed one task per condition, we treat all artist-side quantities as descriptive.

\subsection{Measures and Analytical Approach}
\label{sec:metrics_analysis}
\textbf{Measures.} Participants evaluated each interaction on 7-point Likert scales. Validated instruments included the Single Ease Question (SEQ) \cite{sauro_comparison_2009} for task difficulty, an adapted Psychological Ownership Questionnaire (POQ) \cite{van_dyne_psychological_2004}, and selected Creativity Support Index (CSI) sub-scales \cite{cherry_quantifying_2014}; we added custom metrics for Partnership, Cognitive Support, and confidence change (post-task minus baseline; ``Conf.~$\Delta$'' in Table~\ref{tab:results}), and self-reports of Creative Agency and Semantic Alignment. The principal constructs appear in Table~\ref{tab:results}; Cognitive Support and Partnership were the focal outcomes.

\textbf{Analysis.} Given the ordinal scales and non-normal paired differences (Shapiro--Wilk $p<0.05$ for three of the nine), the primary test was the Wilcoxon Signed-Rank Test on integer item-sum differences, evaluated against its exact signed-rank permutation null (average ranks, zeros dropped; the appropriate small-sample procedure at $N{=}16$). The normal approximation (continuity-corrected) and Pratt's zero-handling serve as sensitivity checks. The nine Wilcoxon tests and the nine LMM System main effects were Benjamini--Hochberg-corrected for the false discovery rate (FDR), the interactions among themselves where cited. Effect size is rank-biserial $r$; reliability was Cronbach's $\alpha$. To corroborate these results, we fit linear mixed-effects models (LMMs) with System Condition, Scenario Type, and their interaction as fixed effects, Block (order) as a covariate, and a participant-level random intercept, estimating the marginal System main effect (averaged across scenarios) and the System $\times$ Scenario interaction.

\section{Results}\label{sec:results}

\subsection{Quantitative Analysis}
\label{sec:quantitative_results}

\begin{figure}[!t]
    \centering
    \includegraphics[width=\linewidth]{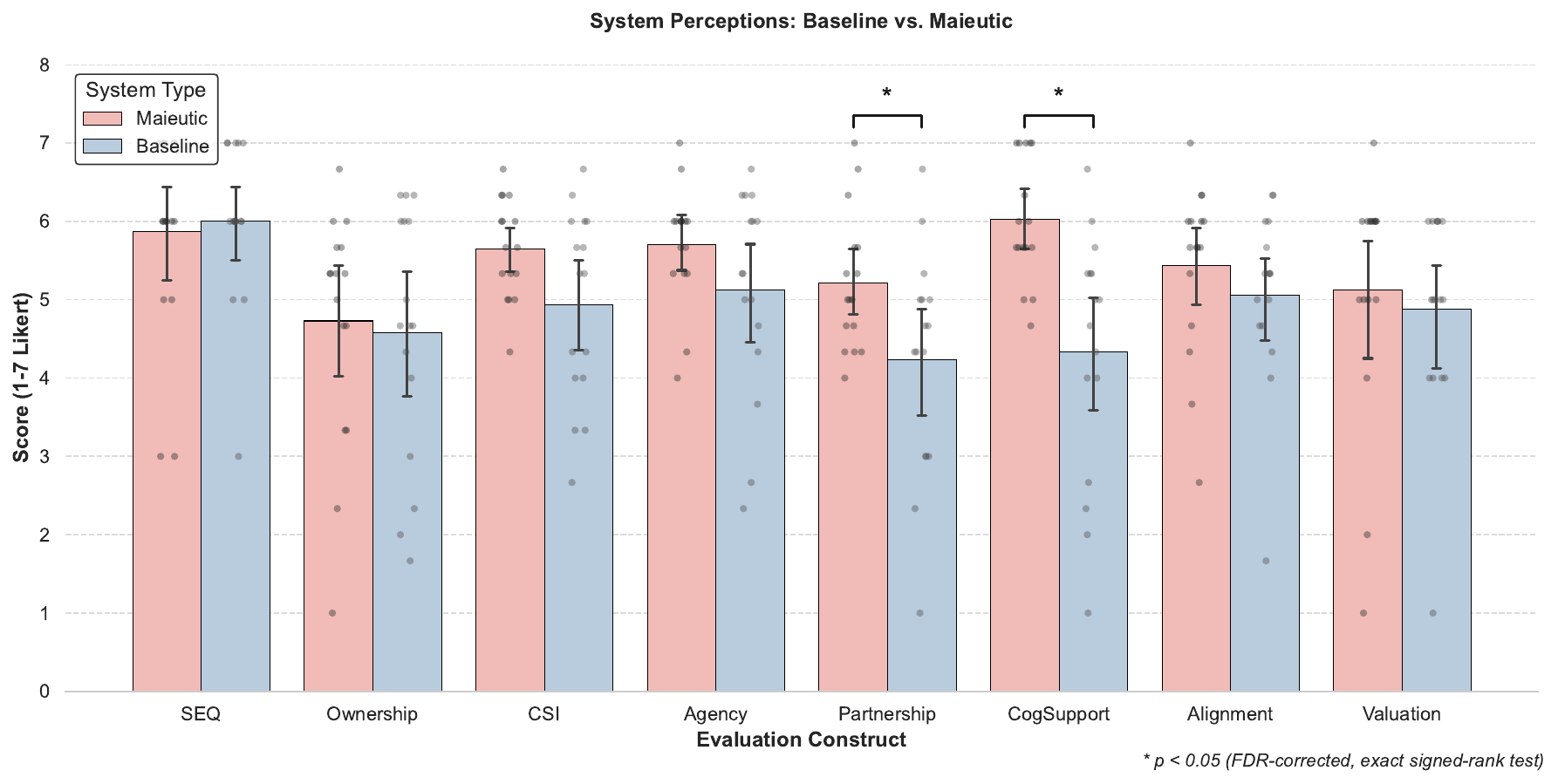}
    \caption{User ratings per construct, Baseline vs.\ Maieutic; error bars: 95\% CIs; asterisks: $p_{\mathrm{FDR}}<0.05$.}
    \label{fig:main_results}
\end{figure}

\begin{table}[!t]
    \centering
    \caption{Effect-size and significance results for the principal constructs ($N{=}16$); Valuation is in the FDR family but not shown. Means and 95\% CIs in Fig.~\ref{fig:main_results}.}
    \label{tab:results}
    \footnotesize
    \begin{tabular}{lcccc}
        \toprule
        Construct & $r$ & $p_{\mathrm{Wil}}^{\mathrm{FDR}}$ & $\beta_{\mathrm{Sys}}$ & $p_{\mathrm{LMM}}^{\mathrm{FDR}}$ \\
        \midrule
        Cog. Support        & 0.971  & \textbf{0.003}* & \textbf{1.69}  & \textbf{$<$0.001}* \\
        Partnership         & 0.752  & \textbf{0.041}* & \textbf{0.98}  & \textbf{0.012}* \\
        Creative Agency     & 0.371  & 0.470           & 0.58           & 0.171 \\
        Sem. Alignment      & 0.352  & 0.470           & 0.38           & 0.424 \\
        CSI                 & 0.610  & 0.138           & 0.71           & 0.098 \\
        Psych. Ownership    & 0.048  & 0.888           & 0.15           & 0.659 \\
        SEQ                 & $-0.250$ & 0.861           & $-0.13$          & 0.659 \\
        Conf. $\Delta$      & 0.154  & 0.779           & 0.19           & 0.659 \\
        \bottomrule
    \end{tabular}
    \begin{flushleft}\vspace{2pt}
    \footnotesize{$r$: rank-biserial effect size (integer item-sum differences); $\beta_{\mathrm{Sys}}$: marginal System effect from the LMM. CSI: Creativity Support Index; SEQ: Single Ease Question. $p_{\mathrm{Wil}}^{\mathrm{FDR}}$: exact signed-rank permutation $p$-values on integer item-sum differences, Benjamini--Hochberg-corrected across the nine study outcomes (the eight shown plus Valuation). Sensitivity analyses using the normal approximation with continuity correction and Pratt zero-handling preserved the focal Cognitive Support result; the Partnership verdict was convention-sensitive. *: $p_{\mathrm{FDR}}<0.05$; \textbf{bold}: focal effect or FDR-significant.}
    \end{flushleft}
\end{table}

\subsubsection{System Main Effects}
Our focal outcome, \textbf{Cognitive Support}, improved significantly under the Maieutic condition (Fig.~\ref{fig:main_results}, Table~\ref{tab:results}), an effect that holds across two complementary analyses (Wilcoxon $p_{\mathrm{FDR}}=0.003$, $r=0.97$; LMM $\beta_{\mathrm{Sys}}=1.69$ (95\% CI $[1.0,2.4]$), $p_{\mathrm{FDR}}<0.001$, marginal $R^2{=}0.40$; residuals normal, Shapiro $p{=}0.07$). Designated a priori as a focal measure with a reliable scale in both conditions (Section~\ref{subsubsec:reliability}), this is not a post-hoc fishing result and is corroborated qualitatively (Section~\ref{sec:qualitative_results}). \textbf{Partnership} improved as well, surviving correction under both the exact signed-rank test ($r=0.75$, $p_{\mathrm{FDR}}=0.041$) and the LMM ($p_{\mathrm{FDR}}=0.012$). Its exact-test verdict is sensitive to the zero-difference convention (Table~\ref{tab:results}), and its Maieutic-side reliability is low, so we treat it as supporting rather than confirmatory evidence (Section~\ref{subsubsec:reliability}). A sign-test sensitivity check, which does not require the signed-rank symmetry assumption, also supported the focal effect but not Partnership. Creativity Support (CSI) trended upward ($\beta{=}0.71$, $p_{\mathrm{LMM}}^{\mathrm{FDR}}{=}0.098$) but did not survive correction.

\subsubsection{Scenario Effects and Interaction}\label{subsubsec:interaction}
The LMMs isolated no marginal Scenario main effects (all $p>0.2$). A System$\times$Scenario interaction trended for Cognitive Support (uncorrected $p=0.030$; BH $q=0.27$ across the nine interactions) but did not survive correction; because the interaction is carried entirely between subjects (cell $n{=}8$; one stimulus per scenario), it is under-powered and specific to these two stimuli. We therefore read the contrast ($+2.67$ Commercial vs.\ $+0.71$ Narrative) only as hypothesis-generating; the direction reverses our expectation.\par A Block (order) effect appeared for Partnership (uncorrected $p{=}0.047$).

\subsubsection{Robustness Checks}\label{subsubsec:reliability}
Exploratory baseline-trait analyses (AI Literacy, Domain Uncertainty, prior commission experience) did not survive multiplicity correction, so we make no demographic-neutrality claim. Internal consistency was acceptable for the focal measure (Cognitive Support $\alpha{=}0.82/0.78$; Baseline/Maieutic), while several secondary constructs were less reliable. An exploratory factor analysis was consistent with the intended Cognitive Support structure, but its small repeated-measures sample precludes strong construct-validity claims.

\subsubsection{Artist-Side Evaluation (Exploratory)}
\label{sec:artist_results}

Table~\ref{tab:artist_results} summarizes the artist-side results. Ratings increased from $B_{\mathrm{base}}$ to $B_{\mathrm{final}}$ on all three reliable dimensions across all eight sampled tasks. Gains were numerically larger under MAIA than under Baseline; all artist-side contrasts are descriptive. Artist autonomy did not change and its scale was unreliable in the final briefs, so we draw no conclusion from it.

\begin{table}[!t]
    \centering
    \caption{Artist-side blind review (descriptive; means over three raters; four paired participants).}
    \label{tab:artist_results}
    \footnotesize
    \setlength{\tabcolsep}{3pt}
    \begin{tabular}{lcccc}
        \toprule
        Measure & Overall gain & MAIA gain & Baseline gain & Paired diff. \\
        \midrule
        VSC               & $+2.76$ & $+3.65$ & $+1.87$ & $+1.78$ \\
        Executability     & $+2.49$ & $+3.19$ & $+1.78$ & $+1.42$ \\
        Professionalism   & $+2.54$ & $+3.58$ & $+1.50$ & $+2.08$ \\
        Autonomy$^\dagger$ & $+0.15$ & $+0.44$ & $-0.14$ & $+0.58$ \\
        \bottomrule
    \end{tabular}
    \begin{flushleft}\vspace{2pt}
    \footnotesize{Artist-side results are descriptive (Section~\ref{sec:artist_results}); `Overall gain' is $B_{\mathrm{base}}\rightarrow B_{\mathrm{final}}$; `Paired diff.' is the participant-paired $\Delta_M{-}\Delta_B$. ICC(2,3)$=0.55$--$0.60$ for VSC, executability, and professionalism. $^\dagger$Autonomy was unreliable ($\alpha{=}{-}0.86$; negative ICC) and is not interpreted.}
    \end{flushleft}
\end{table}

\subsection{Qualitative Analysis}
\label{sec:qualitative_results}
To contextualize these findings, we thematically analyzed open-ended feedback (quotes lightly edited for typographical errors), which traces the mechanisms behind the quantitative effects and surfaces two tensions.

\subsubsection{Cognitive Scaffolding and Drift Control}
Participants experienced the Maieutic condition as instructive, supporting DG1--DG3. Its bounded either/or prompts (DG2) are why it felt more like being \textit{taught} than served: ``It felt like the first system was teaching me how to write a commission rather than just doing it for me,'' and participants valued DG1's ``professional suggestions''. The baseline's unconstrained generation was felt as semantic drift (``it would just generate a bunch of random stuff''), whereas the Maieutic checks made the brief ``much more complete and thorough.''

\subsubsection{Why Ownership Did Not Shift}
Feedback explains the Ownership null (see Discussion): in a transactional commission the brief is a means to the artifact: ``I do not care if it is `mine' or not because I only care about the final result.''

\subsubsection{Tensions: De-visualization and Latency}
Finally, two tensions emerged. First, despite our deliberate text-only design, users conditioned by text-to-image tools craved visuals: ``only text?''; ``showing the final image output might be more helpful.'' Second, the heavy CoT pipeline added latency (``long waiting time''), and some wished to inject details directly, skipping the structured prompts.

\section{Discussion}\label{sec:discussion}

\subsection{Interpreting the Findings}

Across both analytical methods, the full MAIA configuration produced a large, convergent gain in our focal outcome, Cognitive Support (Section~\ref{sec:quantitative_results}). The central inference is that ``requirement discovery'' is a distinct, scaffoldable upstream stage. The mechanism is legible in the qualitative data: participants described the system surfacing dimensions they had not realized were missing (``the first system mentioned and clarified the actual medium and size\ldots this was not a consideration I had''), the traversal of the problem space that laypeople cannot perform alone. This returns to our reading of AI-MC, which assumes a stable intent and optimizes its expression: art commissions instead demand that the intent be constituted, and our results indicate active epistemic co-articulation serves this function.

Semantic Alignment did not differ between conditions (Wilcoxon $p_{\mathrm{FDR}}{=}0.47$; Baseline $M{=}5.06$, Maieutic $M{=}5.44$ on a 7-point scale), and confidence rose in \emph{both} conditions from roughly $3.8$ before the task to about $5.0$ after it ($M{=}5.13$ Maieutic, $4.88$ Baseline). Both patterns are consistent with the scaffolding working as intended: participants rated their final briefs as well aligned with their vision and ended the session more certain that a human artist could execute them. We cannot rule out alternative explanations with these data (for instance, that the briefs are lower-fidelity). The validator gate structurally blocks unratified \textsc{Inferred} content, while qualitative feedback also described less perceived semantic drift. We therefore frame our contribution as a structural mechanism for semantic-drift control (the validator gate): across the 14 Maieutic sessions with retained logs, the gate returned \textsc{Revision-Needed} in 4 sessions (29\%), catching each drift mode the architecture targets (unratified inferences, affective descriptors inflated into visual codes, over-specified \textsc{Discretion} dimensions, and missing JIT definitions). Because the Baseline has no validator by design, drift reduction was not measured as a between-condition outcome. Absolute Cognitive Support under Maieutic was high ($M{=}6.02$ of 7, above the scale midpoint), and participants rated the baseline above it as well ($M{=}4.33$). The effect is a relative improvement, not evidence of ceiling-level quality. The artist-side review provides complementary descriptive evidence on downstream completeness and executability.

The remaining nulls admit honest readings. Psychological Ownership did not shift because, in transactional commissions, the brief is only a means to the artifact; as one participant put it, ownership is irrelevant when ``I only care about the final result'' (Section~\ref{sec:qualitative_results}). Creativity Support (CSI) trended upward ($\beta{=}0.71$, LMM $p_{\mathrm{FDR}}{=}0.098$; uncorrected Wilcoxon $p{=}0.046$), directionally consistent with Cognitive Support but under-powered at $N{=}16$. The Creative Agency and Semantic Alignment nulls likewise reflect uncertainty: with a small sample and borderline Maieutic reliability for Agency ($\alpha{=}0.62$), we cannot claim absence of effect.

\subsection{Design Implications}

Our de-visualization strategy surfaced a tension between epistemic safety and user expectations. Conditioned by text-to-image tools, lay users craved immediate visuals (``only text?''), yet yielding would risk overwriting latent intent with hallucinated imagery, the visual-fixation risk we ground in mental-imagery research \cite{pearson2019human}, which we term \textit{Premature Example Fixation} when triggered by AI-generated examples. We therefore propose \textbf{Phase-Gated Generation}: systems should gate visual outputs until textual Semantic Saturation is reached, so images stay supplementary. This mirrors how professional designers use abstraction to loosen a client's premature outcome fixation before production~\cite{paton2011brief}. An open question is whether such a system should surface artists' reservations about AI-generated reference material before offering it, letting the agent broker a more respectful exchange.

A second implication is methodological. The book-illustration task behaved like the commercial task: both registered as transactional, and Psychological Ownership did not shift, because the simulated, unpaid commission made participants engage the brief more casually than artwork ``for which I am paying.'' Deep emotional projection onto an original character is hard to induce in a constrained lab with convenience-sampled participants, and the interaction decomposition favored the commercial task (an uncorrected interaction), so we treat it as hypothesis-generating. Testing this affective-rich pole will require recruiting, from commission platforms or social media, participants seeking to commission their own established characters.

\subsection{Limitations and Future Work}

Several boundaries qualify these results. First, our artist-side evidence is exploratory: three raters evaluated 16 briefs drawn from 4 of the 16 participants, so the eight task-level units are not independent. Inter-rater agreement was limited, and the condition comparison involved only four paired participants. The task-level rating increase in brief executability and VSC was consistent across all eight sampled tasks, but a full-scale, pre-registered blind review with more artists remains the most important next step. Second, our comparison sets the full configuration against a minimal baseline: the first question for a previously unstudied stage is whether scaffolding helps at all, and the gain could reflect structured process in general rather than maieutic inquiry per se. We therefore frame all claims in terms of the full configuration and leave isolating the sub-agent architecture from the inquiry strategy to follow-up work with intermediate control conditions. Third, the small sample ($N{=}16$) leaves the non-focal nulls under-powered, and carryover cannot be excluded. Partnership is convention-sensitive across the Wilcoxon variants and has low Maieutic-side reliability, so we treat it as supporting evidence rather than a confirmatory result. The System$\times$Scenario interaction did not survive correction. Fourth, all results use a single model (MiniMax M2.5), so cross-model transfer remains open; within that model, the validator also shares the generator's underlying weights, so full validator independence would require a second model. The CoT pipeline added latency that future work must reduce.

\section{Conclusion}

We traced the commissioner's problem to an articulation bottleneck at requirement discovery: the upstream stage where a layperson knows what they feel but must still constitute that intent rather than optimize it. MAIA scaffolds this stage through multi-agent Socratic inquiry under a ``Verification over Invention'' rule. In our within-subjects study ($N{=}16$), the full configuration produced a large gain in Cognitive Support over a minimal baseline, converging across both statistical tests, evidence that requirement discovery is a distinct stage that can be scaffolded. Complementing this, an exploratory artist-side blind review provides descriptive evidence that, in the sampled tasks, the AI-rewritten briefs were rated as more complete and executable by professional artists. We expect this scaffolding principle to transfer to other expert--layperson gaps, for example requirement elicitation in software engineering, wherever a layperson knows what they want but lacks the words to say it.

\bibliographystyle{IEEEtran}
\bibliography{mainref}

@article{camerer:1989:ck,
  author  = {Camerer, Colin and Loewenstein, George and Weber, Martin},
  title   = {The Curse of Knowledge in Economic Settings: An Experimental Analysis},
  doi     = {10.1086/261651},
  journal = {Journal of Political Economy},
  year    = {1989},
  volume  = {97},
  number  = {5},
  pages   = {1232--1254},
}

@article{Hancock_Naaman_Levy_2020, 
    title={{AI}-Mediated Communication: Definition, research agenda, and ethical considerations},
    volume={25}, 
    DOI={10.1093/jcmc/zmz022}, 
    number={1}, 
    journal={Journal of Computer-Mediated Communication}, 
    author={Hancock, Jeffrey T and Naaman, Mor and Levy, Karen}, 
    year={2020}, 
    month={Jan}, 
    pages={89--100}
}

@article{Verheijden_Funk_2023, 
    title={Collaborative diffusion: Boosting designerly co-creation with Generative AI}, 
    DOI={10.1145/3544549.3585680}, 
    journal={Extended Abstracts of the 2023 CHI Conference on Human Factors in Computing Systems}, 
    author={Verheijden, Mathias Peter and Funk, Mathias}, 
    year={2023}, 
    month={Apr}, 
    pages={1–8}
}

@article{Jiang_Wu_Deng_Long_Tang_Li_Liu_Jin_Zhang_Qi_2024, 
    title={Haigen: Towards human-ai collaboration for facilitating creativity and style generation in fashion design}, 
    volume={8}, 
    DOI={10.1145/3678518}, 
    number={3}, 
    journal={Proceedings of the ACM on Interactive, Mobile, Wearable and Ubiquitous Technologies}, 
    author={Jiang, Jianan and Wu, Di and Deng, Hanhui and Long, Yidan and Tang, Wenyi and Li, Xiang and Liu, Can and Jin, Zhanpeng and Zhang, Wenlei and Qi, Tangquan}, 
    year={2024}, 
    month={Aug}, 
    pages={1–27}
}

@article{Fu_Newman_Going_Feng_Lee_2025, 
    title={Exploring the collaborative co-creation process with AI: A case study in novice music production}, 
    DOI={10.1145/3715336.3735829}, 
    journal={Proceedings of the 2025 ACM Designing Interactive Systems Conference}, 
    author={Fu, Yue and Newman, Michele and Going, Lewis and Feng, Qiuzi and Lee, Jin Ha}, 
    year={2025}, 
    month={Jul}, 
    pages={1298–1312}
}

@inproceedings{Inie_Falk_Tanimoto_2023, series={CHI ’23}, title={Designing Participatory {AI}: Creative Professionals’ Worries and Expectations about Generative {AI}}, url={http://dx.doi.org/10.1145/3544549.3585657}, DOI={10.1145/3544549.3585657}, booktitle={Extended Abstracts of the 2023 CHI Conference on Human Factors in Computing Systems}, publisher={ACM}, author={Inie, Nanna and Falk, Jeanette and Tanimoto, Steve}, year={2023}, month = apr, pages={1--8}, collection={CHI ’23},
  address = {New York, NY, USA}
}

@article{Shi_Gao_Jiao_Cao_2023, 
    title={Understanding design collaboration between designers and Artificial Intelligence: A systematic literature review}, 
    volume={7}, 
    DOI={10.1145/3610217}, 
    number={CSCW2}, 
    journal={Proceedings of the ACM on Human-Computer Interaction}, 
    author={Shi, Yang and Gao, Tian and Jiao, Xiaohan and Cao, Nan}, 
    year={2023}, 
    month={Sep}, 
    pages={1–35}
}

@misc{Mieczkowski_Hancock_2022, title={Examining Agency, Expertise, and Roles of {AI} Systems in {AI}-Mediated Communication}, url={http://dx.doi.org/10.31219/osf.io/asnv4}, DOI={10.31219/osf.io/asnv4}, publisher={Center for Open Science}, author={Mieczkowski, Hannah and Hancock, Jeffrey}, year={2022}, month = jul, howpublished={Open Science Framework}, note={Preprint} }

@article{Mieczkowski_Hancock_Naaman_Jung_Hohenstein_2021, title={{AI}-Mediated Communication: Language Use and Interpersonal Effects in a Referential Communication Task}, volume={5}, ISSN={2573-0142}, url={http://dx.doi.org/10.1145/3449091}, DOI={10.1145/3449091}, number={CSCW1}, journal={Proceedings of the ACM on Human-Computer Interaction}, publisher={Association for Computing Machinery (ACM)}, author={Mieczkowski, Hannah and Hancock, Jeffrey T. and Naaman, Mor and Jung, Malte and Hohenstein, Jess}, year={2021}, month = apr, pages={1--14} }

@article{Meng_Zhang_Qin_Lee_Lee_2025, 
    title={AI-mediated social support: The prospect of human–ai collaboration}, 
    volume={30}, 
    DOI={10.1093/jcmc/zmaf013}, 
    number={4}, 
    journal={Journal of Computer-Mediated Communication}, 
    author={Meng, Jingbo and Zhang, Renwen and Qin, Jiaqi and Lee, Yu-Jen and Lee, Yi-Chieh}, 
    year={2025}, 
    month={May}
}

@article{fang_generative_2025,
	title = {Generative {AI}-enhanced human-{AI} collaborative conceptual design: {A} systematic literature review},
	volume = {97},
	issn = {0142694X},
	shorttitle = {Generative {AI}-enhanced human-{AI} collaborative conceptual design},
	doi = {10.1016/j.destud.2025.101300},
	language = {en},
	urldate = {2025-09-09},
	journal = {Design Studies},
	author = {Fang, Cong and Zhu, Yujie and Fang, Le and Long, Yonghao and Lin, Huan and Cong, Yangfan and Wang, Stephen Jia},
	month = mar,
	year = {2025},
	pages = {101300},
}

@inproceedings{huang_causalmapper_2023,
	address = {Virtual Event USA},
	title = {{CausalMapper}: {Challenging} designers to think in systems with {Causal} {Maps} and {Large} {Language} {Model}},
	isbn = {979-8-4007-0180-1},
	shorttitle = {{CausalMapper}},
	doi = {10.1145/3591196.3596818},
	language = {en},
	urldate = {2025-09-09},
	booktitle = {Creativity and {Cognition}},
	publisher = {ACM},
	author = {Huang, Ziheng and Quan, Kexin and Chan, Joel and MacNeil, Stephen},
	month = jun,
	year = {2023},
	pages = {325--329},
}

@article{valk_ideation_2023,
	title = {The {Ideation} {Compass}: supporting interdisciplinary creative dialogues with real time visualization},
	volume = {11},
	issn = {2165-0349, 2165-0357},
	shorttitle = {The {Ideation} {Compass}},
	doi = {10.1080/21650349.2022.2142674},
	language = {en},
	number = {2},
	urldate = {2025-09-09},
	journal = {International Journal of Design Creativity and Innovation},
	author = {Välk, Sander and Thabsuwan, Chitipat and Mougenot, Céline},
	month = apr,
	year = {2023},
	pages = {99--116},
}

@misc{chang_framework_2024,
	title = {A {Framework} for {Collaborating} a {Large} {Language} {Model} {Tool} in {Brainstorming} for {Triggering} {Creative} {Thoughts}},
	doi = {10.48550/arXiv.2410.11877},
	language = {en},
	urldate = {2025-09-10},
	publisher = {arXiv},
	author = {Chang, Hung-Fu and Li, Tong},
	month = oct,
	year = {2024},
	note = {arXiv:2410.11877 [cs]},
}

@article{jansson_design_1991,
	title = {Design fixation},
	volume = {12},
	issn = {0142-694X},
	doi = {10.1016/0142-694X(91)90003-F},
	number = {1},
	journal = {Design Studies},
	author = {Jansson, David G. and Smith, Steven M.},
	year = {1991},
	pages = {3--11},
}

@article{belkin1980anomalous,
  title={Anomalous states of knowledge as a basis for information retrieval},
  author={Belkin, Nicholas J},
  journal={Canadian journal of information science},
  volume={5},
  number={1},
  pages={133--143},
  year={1980}
}

@article{eppler2007knowledge,
  title={Knowledge communication problems between experts and decision makers: An overview and classification},
  author={Eppler, Martin J},
  journal={Electronic Journal of Knowledge Management},
  volume={5},
  number={3},
  pages={pp291--300},
  year={2007}
}

@article{eppler2004overload,
  author  = {Eppler, Martin J. and Mengis, Jeanne},
  title   = {The Concept of Information Overload: A Review of Literature from Organization Science, Accounting, Marketing, MIS, and Related Disciplines},
  journal = {The Information Society},
  year    = {2004},
  volume  = {20},
  number  = {5},
  pages   = {325--344},
  doi     = {10.1080/01972240490507974},
  publisher = {Taylor \& Francis}
}

@article{cao2020expertise,
  title={Expertise style transfer: A new task towards better communication between experts and laymen},
  doi={10.48550/arXiv.2005.00701},
  author={Cao, Yixin and Shui, Ruihao and Pan, Liangming and Kan, Min-Yen and Liu, Zhiyuan and Chua, Tat-Seng},
  journal={arXiv preprint arXiv:2005.00701},
  year={2020}
}

@inproceedings{liu2025exploring, series={CHI ’25}, title={Exploring the Design Space of Real-time {LLM} Knowledge Support Systems: A Case Study of Jargon Explanations}, url={http://dx.doi.org/10.1145/3706598.3714262}, DOI={10.1145/3706598.3714262}, booktitle={Proceedings of the 2025 CHI Conference on Human Factors in Computing Systems}, publisher={ACM}, author={Liu, Yuhan and Shah, Aadit and Ackerman, Jordan and Saha, Manaswi}, year={2025}, month = apr, pages={1--20}, collection={CHI ’25},
  address = {New York, NY, USA}
}

@article{bischof2011caring,
  title={Caring for Clarity in Knowledge Communication.},
  author={Bischof, Nicole and Eppler, Martin J},
  journal={J. Univers. Comput. Sci.},
  volume={17},
  number={10},
  pages={1455--1473},
  year={2011}
}

@article{song2025personalized,
  title={Personalized Real-time Jargon Support for Online Meetings},
  doi={10.48550/arXiv.2508.10239},
  author={Song, Yifan and Au, Wing Yee and Wong, Hon Yung and Bailey, Brian P and August, Tal},
  journal={arXiv preprint arXiv:2508.10239},
  year={2025}
}

@article{pearson2019human,
  title={The human imagination: the cognitive neuroscience of visual mental imagery},
  doi={10.1038/s41583-019-0202-9},
  author={Pearson, Joel},
  journal={Nature reviews neuroscience},
  volume={20},
  number={10},
  pages={624--634},
  year={2019},
  publisher={Nature Publishing Group UK London}
}

@article{tversky1974judgment,
  title={Judgment under Uncertainty: Heuristics and Biases: Biases in judgments reveal some heuristics of thinking under uncertainty.},
  doi={10.1126/science.185.4157.1124},
  author={Tversky, Amos and Kahneman, Daniel},
  journal={science},
  volume={185},
  number={4157},
  pages={1124--1131},
  year={1974},
  publisher={American association for the advancement of science}
}

@book{csikszentmihalyi1990flow,
  title={Flow: The psychology of optimal experience},
  author={Csikszentmihalyi, Mihaly},
  volume={1990},
  year={1990},
  publisher={Harper \& Row New York}
}

@article{shneiderman2007creativity,
  title={Creativity support tools: accelerating discovery and innovation},
  doi={10.1145/1323688.1323689},
  author={Shneiderman, Ben},
  journal={Communications of the ACM},
  volume={50},
  number={12},
  pages={20--32},
  year={2007},
  publisher={ACM New York, NY, USA}
}

@inproceedings{sauro_comparison_2009,
	address = {Boston MA USA},
	title = {Comparison of three one-question, post-task usability questionnaires},
	isbn = {978-1-60558-246-7},
	doi = {10.1145/1518701.1518946},
	language = {en},
	urldate = {2026-02-05},
	booktitle = {Proceedings of the {SIGCHI} {Conference} on {Human} {Factors} in {Computing} {Systems}},
	publisher = {ACM},
	author = {Sauro, Jeff and Dumas, Joseph S.},
	month = apr,
	year = {2009},
	pages = {1599--1608}
}

@article{van_dyne_psychological_2004,
	title = {Psychological ownership and feelings of possession: three field studies predicting employee attitudes and organizational citizenship behavior},
	volume = {25},
	copyright = {http://onlinelibrary.wiley.com/termsAndConditions\#vor},
	issn = {0894-3796, 1099-1379},
	shorttitle = {Psychological ownership and feelings of possession},
	doi = {10.1002/job.249},
	language = {en},
	number = {4},
	urldate = {2026-02-05},
	journal = {Journal of Organizational Behavior},
	author = {Van Dyne, Linn and Pierce, Jon L.},
	month = jun,
	year = {2004},
	pages = {439--459}
}

@article{cherry_quantifying_2014,
	title = {Quantifying the {Creativity} {Support} of {Digital} {Tools} through the {Creativity} {Support} {Index}},
	volume = {21},
	issn = {1073-0516, 1557-7325},
	doi = {10.1145/2617588},
	language = {en},
	number = {4},
	urldate = {2026-02-05},
	journal = {ACM Transactions on Computer-Human Interaction},
	author = {Cherry, Erin and Latulipe, Celine},
	month = aug,
	year = {2014},
	pages = {1--25},
}

@article{paton2011brief, title={Briefing and reframing: A situated practice}, volume={32}, ISSN={0142-694X}, url={http://dx.doi.org/10.1016/j.destud.2011.07.002}, DOI={10.1016/j.destud.2011.07.002}, number={6}, journal={Design Studies}, publisher={Elsevier BV}, author={Paton, Bec and Dorst, Kees}, year={2011}, month=nov, pages={573--587} }

@inproceedings{jung_maieutic_2022,
  title     = {Maieutic Prompting: Logically Consistent Reasoning with Recursive Explanations},
  doi       = {10.18653/v1/2022.emnlp-main.82},
  author    = {Jung, Jaehun and Qin, Lianhui and Welleck, Sean and Brahman, Faeze and Bhagavatula, Chandra and Le Bras, Ronan and Choi, Yejin},
  booktitle = {Proceedings of the 2022 Conference on Empirical Methods in Natural Language Processing (EMNLP)},
  pages     = {1116--1137},
  year      = {2022},
  publisher = {Association for Computational Linguistics}
}

@book{stokes_creativity_2006,
  author    = {Stokes, Patricia D.},
  title     = {Creativity from Constraints: The Psychology of Breakthrough},
  year      = {2006},
  publisher = {Springer Publishing Company},
  address   = {New York}
}

@article{slovic_construction_1995, title={The construction of preference.}, volume={50}, ISSN={0003-066X}, url={http://dx.doi.org/10.1037/0003-066X.50.5.364}, DOI={10.1037/0003-066x.50.5.364}, number={5}, journal={American Psychologist}, publisher={American Psychological Association (APA)}, author={Slovic, Paul}, year={1995}, month=may, pages={364--371} }

\end{document}